\documentclass[sigconf,screen]{acmart} 
\AtBeginDocument{%
  \providecommand\BibTeX{{%
    \normalfont B\kern-0.5em{\scshape i\kern-0.25em b}\kern-0.8em\TeX}}}

\setcopyright{acmlicensed}
\acmPrice{15.00}
\acmDOI{10.1145/3611643.3616278}
\acmYear{2023}
\copyrightyear{2023}
\acmSubmissionID{fse23main-p314-p}
\acmISBN{979-8-4007-0327-0/23/12}
\acmConference[ESEC/FSE '23]{Proceedings of the 31st ACM Joint European Software Engineering Conference and Symposium on the Foundations of Software Engineering}{December 3--9, 2023}{San Francisco, CA, USA}
\acmBooktitle{Proceedings of the 31st ACM Joint European Software Engineering Conference and Symposium on the Foundations of Software Engineering (ESEC/FSE '23), December 3--9, 2023, San Francisco, CA, USA}
\received{2023-02-02}
\received[accepted]{2023-07-27}

\usepackage{color}

\definecolor{grayblue}{rgb}{0.2,0.29,0.79}
\definecolor{darkgreen}{rgb}{0.2,0.55,0.1}
\definecolor{violet}{rgb}{0.54,0.17,0.88}

\newcommand{\myadd}[1]
{
   {\noindent\color{red}\bf #1}
}
\newcommand{\mydel}[1]
{
   {\noindent\color{blue} \sout{#1}}
}

\newcommand{\ma}[1]
{
   {\noindent\color{red}\bf [#1]$_{\scriptscriptstyle\textit{ma}}$}
}

\newcommand{\feng}[1]
{
   {\noindent\color{red}\bf [#1]$_{\scriptscriptstyle\textit{feng}}$}
}
\newcommand{\gao}[1]
{
   {\noindent\color{grayblue}\bf [#1]$_{\scriptscriptstyle\textit{gao}}$}
}

\newcommand{\wang}[1]
{
   {\noindent\color{violet}\bf [#1]$_{\scriptscriptstyle\textit{wang}}$}
}

\newcommand{\includeAuthorComments}[1]
{
   \ifthenelse{\equal{#1}{0}}
   {
      \renewcommand{\feng}[1]
      {
         {} 
      }
      \renewcommand{\gao}[1]
      {
         {} 
      }
      \renewcommand{\wang}[1]
      {
         {} 
      }
       \renewcommand{\ma}[1]
      {
         {} 
      }
        \renewcommand{\myadd}[1]
      {
         {} 
      }
        \renewcommand{\mydel}[1]
      {
         {} 
      }
      
   }{}
}

\newcommand{\rqOne}{\textbf{RQ1. How do AI-enabled MSF-based perception systems perform against common corrupted signals?}}
\newcommand{\rqTwo}{\textbf{RQ2. How sensitive is AI-enabled MSF when facing spatial and temporal misalignment of sensors?}}
\newcommand{\rqThree}{\textbf{RQ3. To what extent are existing sensing components coupled of an AI-enabled MSF system?
}}

\newcommand{\rqFour}{\textbf{RQ4. 
What is the weakness of different AI-enabled MSF mechanisms and is it possible to repair them?
}} 

\newcommand{\website}{https://sites.google.com/view/ai-msf-benchmark}

\usepackage{graphicx}
\usepackage{xcolor}
\graphicspath{{figs/}}
\usepackage{makecell}
\usepackage{tcolorbox}
\tcbuselibrary{skins, breakable, theorems}
\usepackage{diagbox}
\usepackage{hyperref}
\usepackage{caption}
\usepackage{multirow}
\usepackage{subfigure}
\usepackage{threeparttable}
\makeatletter  
\newif\if@restonecol  
\makeatother  
\usepackage[linesnumbered,ruled,vlined]{algorithm2e}
\usepackage{algpseudocode}  
\usepackage{enumitem}
\usepackage{calc}
\usepackage{booktabs}
\usepackage{colortbl}

\definecolor{lightgray}{HTML}{eeeeee}
\definecolor{tab_red}{rgb}{1,0.76,0.71}

\newcommand{\midsepremove}{\aboverulesep = 0mm \belowrulesep = 0mm}
    \midsepremove
    \newcommand{\midsepdefault}{\aboverulesep = 0mm \belowrulesep = 0mm}
    \midsepdefault

\begin{document}



\title{Benchmarking Robustness of AI-Enabled Multi-sensor Fusion Systems: Challenges and Opportunities}


\author{Xinyu Gao}
\affiliation{
  \institution{State Key Laboratory for Novel Software Technology}
\country{Nanjing University, China}
}

\author{Zhijie Wang}
\affiliation{
  \institution{University of Alberta}
  \city{Edmonton}
\country{Canada}
}

\author{Yang Feng}
\authornote{Yang Feng and Lei Ma are the corresponding authors.}
\affiliation{
  \institution{State Key Laboratory for Novel Software Technology}
\country{Nanjing University, China}
}

\author{Lei Ma}
\authornotemark[1]
\affiliation{%
  \institution{The University of Tokyo, Japan}
  \city{}
  \state{}
  \country{University of Alberta, Canada}
  }

\author{Zhenyu Chen}
\affiliation{
  \institution{State Key Laboratory for Novel Software Technology}
\country{Nanjing University, China}
}

\author{Baowen Xu}
\affiliation{
  \institution{State Key Laboratory for Novel Software Technology}
\country{Nanjing University, China}
}


\begin{abstract}
Multi-Sensor Fusion (MSF) based perception systems have been the foundation in supporting many industrial applications and domains, such as self-driving cars, robotic arms, and unmanned aerial vehicles.
Over the past few years, the fast progress in data-driven artificial intelligence (AI) has brought a fast-increasing trend to empower MSF systems by deep learning techniques to further improve performance, especially on intelligent systems and their perception systems. 
Although quite a few AI-enabled MSF perception systems and techniques have been proposed, up to the present, limited benchmarks that focus on MSF perception are publicly available.
Given that many intelligent systems such as self-driving cars are operated in safety-critical contexts where perception systems play an important role, there comes an urgent need for a more in-depth understanding of the performance and reliability of these MSF systems. 

To bridge this gap, we initiate an early step in this direction and construct a public benchmark of AI-enabled MSF-based perception systems including three commonly adopted tasks (i.e., object detection, object tracking, and depth completion). 
Based on this, to comprehensively understand MSF systems' robustness and reliability, we design 14 common and realistic corruption patterns to synthesize large-scale corrupted datasets. We further perform a systematic evaluation of these systems through our large-scale evaluation and identify the following key findings: (1) existing AI-enabled MSF systems are not robust enough against corrupted sensor signals; (2) small synchronization and calibration errors can lead to a crash of AI-enabled MSF systems; (3) existing AI-enabled MSF systems are usually tightly-coupled in which bugs/errors from an individual sensor could result in a system crash; (4) the robustness of MSF systems can be enhanced by improving fusion mechanisms.
Our results reveal the vulnerability of the current AI-enabled MSF perception systems, calling for researchers and practitioners to take robustness and reliability into account when designing AI-enabled MSF. Our benchmark, code, and detailed evaluation results are publicly available at \href{\website}{\website}.
\end{abstract}




\maketitle

\section{Introduction}
Multi-sensor fusion (MSF) refers to the technique that combines data from multiple sources of sensors to achieve specific tasks, which has been widely adopted in many real-world complex systems.
The integration of information from different sensors avoids the inherent perception limitations of individual sensors and improves the system's overall performance.
Over the past years, MSF-based perception systems have been widely used in various industrial domains and safety-critical applications, such as self-driving cars~\cite{kato2018autoware}, unmanned aerial vehicles~\cite{samaras2019deep}, and robotic systems~\cite{mao2019brain}. 

With recent advances in data-driven artificial intelligence (AI), there comes an increasing trend in proposing deep learning (DL) techniques to further enable more advanced heterogeneous data processing from different sensors, in order to achieve more accurate perception and prediction.
Given the advantage of deep neural networks (DNNs) in processing and extracting complex semantic information from sensors' data (e.g., image, point cloud), AI-enabled MSF has been increasingly adopted in the perception systems of autonomous driving~\cite{kato2018autoware,samaras2019deep}.

The rapid development of AI-enabled MSF systems also brings challenges and concerns. One of the biggest concerns is the lack of a deep understanding of the current AI-enabled MSF's reliability. In practice, an AI-enabled MSF system could behave incorrectly and lead to severe accidents in safety-critical contexts, especially in autonomous driving~\cite{GoogleAccident, TeslaAccident2}. Thus, it is highly desirable to enable testing, analysis, and systematic assessment of such intelligent systems beforehand comprehensively. One common practice to enable such quality assurance activities in AI/SE communities is to establish a benchmark that enables both researchers and practitioners to perform systematic studies and develop novel techniques to better fulfill important quality requirements.
However, to the best of our knowledge, up to the present, few benchmarks specifically designed for AI-enabled MSF are yet available. It is unclear whether and to what extent the potential quality issues and risks can be, how they are brought from each sensing unit, and their impacts on the integration and the state-of-the-art information fusion processes.

\begin{figure*}
    \centering
    \includegraphics[width = 0.95\linewidth]{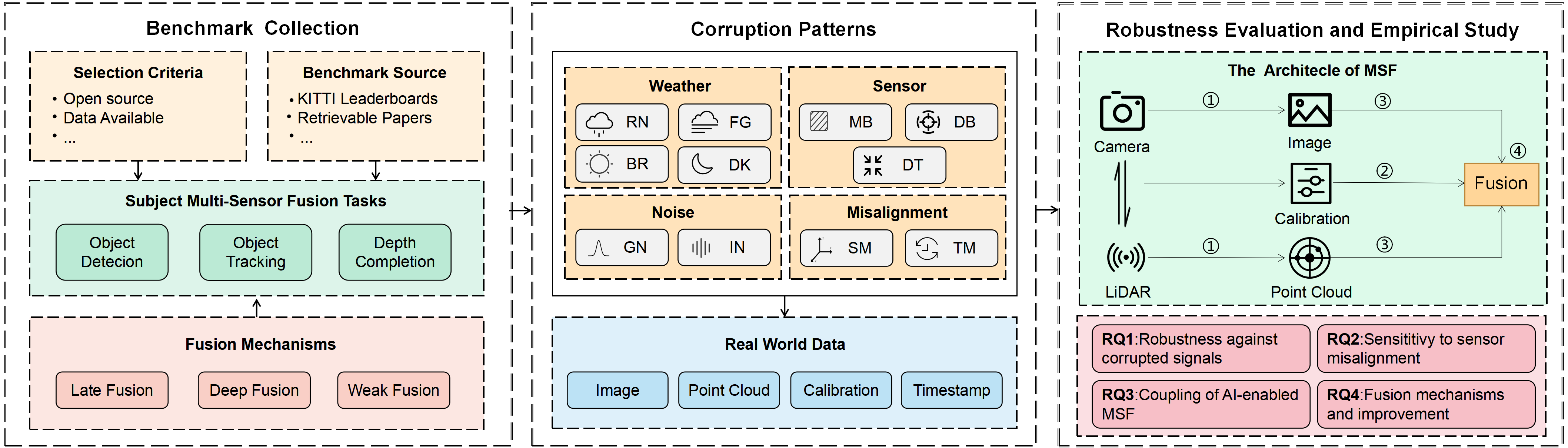}
    \caption{Workflow summary of AI-enabled MSF benchmark construction, and high-level empirical study design.}
    \label{fig: architecture}
\end{figure*}

To bridge this gap, in this paper, we initiate an early step to present a benchmark and perform an empirical study of AI-enabled MSF perception systems. Fig.~\ref{fig: architecture} summarizes the high-level design and workflow of our benchmark construction and our empirical study, in which we mainly investigate the following research questions, aiming to identify the potential challenges and opportunities:

\begin{itemize}[leftmargin=*]

\item {\rqOne} This RQ aims to investigate the potential risks of AI-enabled MSF systems against corrupted signals that commonly occur in the operational environments. Through a large-scale evaluation on eleven types of corrupted sensor signals, we find that the current AI-enabled MSF systems are not robust enough, especially against weather condition changes.

\item {\rqTwo} In the practical open and wild environment, it is almost impossible to always maintain perfect calibration or precise time synchronization of the system across sensors. RQ2 aims to investigate the sensitivity of AI-enabled MSF to spatial and temporal misalignment. Our experiment results reveal that even small calibration or synchronization issues could lead to abnormal behaviors of the system.

\item {\rqThree} A robust and reliable MSF should not completely fail when one or a part of the whole sensing modules lose the source signal. RQ3 aims to investigate how AI-enabled MSF systems can be impacted when one source of the signal is partially/completely lost. Overall, we find that the tightly-coupled architecture of AI-enabled MSF systems exhibits less robustness against signal loss.

\item {\rqFour} RQ4 aims to investigate the unique advantages of each fusion mechanism and potential opportunities for improving the robustness of
AI-enabled 
MSF systems. Our results demonstrate that deep fusion is more robust in some cases, however, weak and late fusion can be easier to be repaired in terms of robustness against corruption patterns.
\end{itemize}

To sum up, this work makes the following contributions:
\begin{itemize}[leftmargin=*]
    \item \textbf{Benchmark}. We initiate to create an early public benchmark of AI-enabled MSF-based perception systems. This provides a common ground for the study and analysis of AI-enabled MSF systems' robustness and enables future quality assurance research in this direction.
    \item \textbf{Empirical Study}. Based on the benchmark, we perform a large-scale empirical study of AI-enabled MSF systems to investigate their current status regarding robustness.
    \item \textbf{Discussion}. We further make discussions about existing AI-enabled MSF systems and future directions, including the unique advantages of different fusion mechanisms as well as the opportunities of their robustness enhancement.
\end{itemize}

To the best of our knowledge, this paper is among the very early research to benchmark and investigate the MSF system, which is a common and representative AI system composed of multiple sensing channels and corresponding models.
On one hand, at present, it is not clear how much and to what extent each sensing unit could impact the integrated sensing results of an MSF; it is not clear how the issues of different sensing units and channels are involved and propagate to the final results of different MSF designs either. Creating a benchmark at the current stage enables to investigate these important questions quantitatively, which also enables further relevant quality assurance research along this direction.
On the other hand, in general, MSF-based perception systems play a key role to enable autonomous and intelligent systems, which potentially has a big impact on many applications and domains. With the recent fast pace in transforming into the data-driven intelligent era, we believe an early stage benchmark and investigation of the current MSF systems empowered by deep learning would also benefit the practitioners in understanding the limitation and proposing better MSF engineering techniques, paving the path towards designing safe and reliable autonomous intelligent systems.

\section{Background}
\label{sec: fusion mechanisms}

\subsection{Perception Systems in Intelligent Systems}

An intelligent system (e.g., a self-driving car, a robotic, an unmanned aerial vehicle) is usually a complex system composed of various subsystems. These subsystems are cooperated to ensure safe and reliable operations of the intelligent system.
The perception system is one of the key components in an intelligent system, which is in charge of sensing and processing environmental information through sensors to perform crucial tasks, e.g., object detection and object tracking. The prediction results from perception systems are then propagated to other components in the intelligent system such as planning and control systems.
The perception systems lay the foundation of the intelligent system’s workflow in perceiving and understanding the environment, which also significantly impact the quality and reliability of the whole system. 

Most industrial-level systems leverage multi-sensor fusion (MSF) strategy to avoid inherent perception limitations of individual sensors and thus sense the environment more reliably~\cite{peng2020apollo}. 
For instance, the camera and LiDAR are usually fused in self-driving cars since camera is more effective in capturing semantic information and LiDAR could provide more accurate geographic information~\cite{survey_cui2021deep}.
As shown in Fig.~\ref{fig: architecture} (right part), each sensor in a camera-LiDAR fusion first senses the surrounding environment individually.
Then, signals from different sensors are transformed into the same coordinate system and matched across the timestamps based on the temporal and spatial calibration among sensors.
Finally, the fusion module receives calibrated and synchronized signals from different sensors, and fuses them to make predictions for downstream tasks.

\subsection{AI-enabled Multi-Sensor Fusion}
Different from traditional MSF that only fuses the data or output, AI-enabled MSF also has the possibility to fuse the deep semantic features learned by DNNs. We take the fusion of camera and LiDAR as an example (right part of Fig.~\ref{fig: architecture}) in the following sections.

Each branch that processes signals in AI-enabled MSF can be represented as a composite function chain (Eq.~\ref{eq:msf}) that maps the input data $M$ to the output result $f^L$.
\begin{equation}
\small
    \label{eq:msf}
      f^{L} = F^{(L)}(F^{(L-1)}(\cdots(F^{(0)}(M)))),
\end{equation}
where $L$ denotes the depth of a branch.
The medium output in the chain, i.e., $F^{(j)}(\cdot), j = \{1,2,\ldots,L-1\}$, represents the output from $j$th hidden layer in a DNN.

\begin{figure}[t]
    \centering
    \includegraphics[width = 0.9\linewidth]{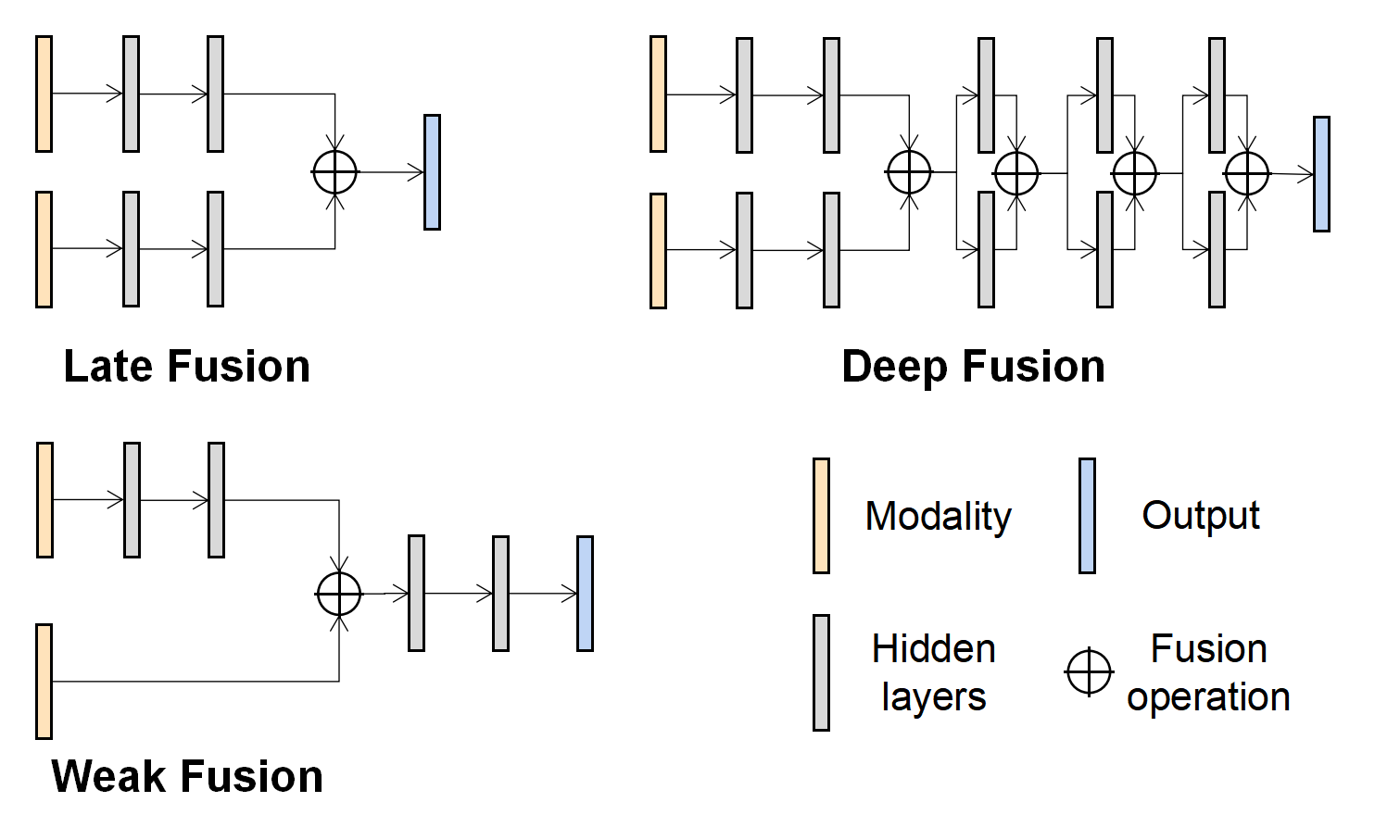}
    \caption{Different AI-enabled MSF mechanisms.}
    \label{fig: fusion}
\end{figure}

Based on the stage where the fusion is made~\cite{survey_feng2020deep, survey_huang2022multi} (see Fig.~\ref{fig: fusion}), AI-enabled MSF can be categorized into four different mechanisms at a high level: \textit{early fusion}, \textit{late fusion}, \textit{deep fusion}, and \textit{weak fusion}. Since \textit{early fusion} is not commonly used in AI-enabled MSF, we focus on the other three fusion mechanisms in the rest of this paper.
For a $L$ layer deep neural network, we denote $M_i$ and $M_j$ as two different modalities and define $\oplus$ as a fusion operation. Now we briefly introduce each MSF mechanism.

\textbf{Late fusion} directly combines the output results of each branch, which can be formulated as:
\begin{equation}
\small
\begin{aligned}
     f^{L} 
     &= F^{(L)}(F_{i}^{(L-1)}(
     \cdots(F_{i}^{(0)}(M_{i}))))\\
     &\oplus
     F^{(L)}(F_{j}^{(L-1)}(
     \cdots(F_{j}^{(0)}(M_{j} ))
     ))
\end{aligned}     
\end{equation}

Each branch in late fusion process data from sensors independently and does not depend on specific network architecture. Compared with other fusion mechanisms, late fusion is highly flexible. For instance, late fusion can easily combine image-based object detectors and LiDAR-based ones. Late fusion does not involve hidden feature interaction, which also leads to higher efficiency.

\textbf{Deep fusion} involves frequent interactions among hidden features from different branches to gain rich semantic information. Suppose that the depth of branch $i$ is greater than that of branch $j$,
when only one feature fusion is performed, the deep fusion can be formulated as:
\begin{equation}
\small
\begin{aligned}
      f^{L} &= F^{(L)}(\cdots(F^{(L^{*}_{i}+1)}
      (
      F^{(L_{i}^*)}(\cdots(F^{(0)}(M_{i}))) \\
      &\oplus
      F^{(L_{j}^*)}(\cdots(F^{(0)}(M_{j})))
      )
      )),
\end{aligned}     
\end{equation}
where $L_{i}^*$, $L_{j}^*$ denotes that the fusion starts from $i$th and $j$th hidden layers, respectively.


\textbf{Weak fusion} does not fuse the hidden features nor fuse the output results. Instead, weak fusion adopts rule-based methods to transform data from one branch to guide the process of data in another branch. The process of weak fusion can be described as:
\begin{equation}
\small
     f^{L} = F^{(L)}(F^{(L-1)}(
     \cdots F^{(0)}(
     G(M_{i}) 
     \oplus M_{j}
     ))),
\end{equation}
where G is the function that extracts the guidance from branch $i$.
One typical example of weak fusion is extracting the frustums in the point cloud data using the 2D detection bounding boxes from the image as guidance~\cite{qi2018frustum,wang2019frustum}.


\section{Benchmark Construction}
\label{sec: benchmark}

\subsection{Benchmark Collection}
To collect as many appropriate AI-enabled MSF perception systems as possible for our study, we mainly focus on two sources: (1) the leaderboard of KITTI benchmark~\cite{kitti_geiger2012we}, and (2) existing MSF-related literature. KITTI is a public autonomous driving benchmark that involves several different perception tasks.
For MSF-related literature, we collect papers published in relevant top-tier conferences and journals during the last four years, covering software engineering, robotics, computer vision, etc.
We refer readers to our supplementary website~\cite{website} for a complete list of selected venues. 
Eventually, we selected 7 state-of-the-art MSF systems from these two sources based on the following criteria:
\begin{itemize}[leftmargin=*]
\item \textbf{Multi-sensors}. An MSF system should involve two or more types of different sensors.
\item \textbf{Open-source}. An MSF system should be open-source so that we can conduct experimental evaluations and enable further replication studies.
\item \textbf{Data available}. An MSF system should have open-source data for training and evaluation.
\item \textbf{Representative task}. An MSF system should be designed for representative perception tasks with real-world applications, e.g., object detection.
\end{itemize}

\begin{table}[t]
\renewcommand{\arraystretch}{1.2}
\setlength{\tabcolsep}{3.6pt}
\footnotesize
\centering
\caption{The collected MSF systems. Performance of each system is evaluated by task-specific metrics (detailed in Sec.~\ref{sec: eval metrics}).}
\label{tab: fusion systems}
\begin{tabular}{lccccc}
\toprule
\textbf{System}                    & \textbf{Task}    & \textbf{Fusion} & \textbf{Year} & \textbf{Modality} & \textbf{Performance} \\
\midrule
EPNet~\cite{huang2020epnet}        & Object Detection & Deep            & 2020          & C+L             & 82.70                \\
FConv~\cite{wang2019frustum}       & Object Detection & Weak            & 2019          & C+L             & 79.06                \\
CLOCs~\cite{pang2020clocs}         & Object Detection & Late            & 2020          & C+L             & 79.70                \\
JMODT~\cite{huang2021jmodt}        & Object Tracking  & Deep            & 2021          & C+L             & 86.12                \\
DFMOT~\cite{wang2022deepfusionmot} & Object Tracking  & Late            & 2022          & C+L             & 80.17                \\
TWISE~\cite{imran2021twise}        & Depth Completion & Deep            & 2021          & C+L             & 1009.64              \\
MDANet~\cite{ke2021mdanet}         & Depth Completion & Deep            & 2021          & C+L             & 898.39              
\\\bottomrule
\end{tabular}
\end{table}

Table~\ref{tab: fusion systems} summarizes the seven MSF systems selected in our benchmark.
These seven systems cover three different tasks and three different fusion mechanisms. Due to the page limit, we refer audiences to our supplementary website~\cite{website} for details of each MSF system.

\subsection{Corruption Patterns}

Operational environments of many MSF systems are usually open with unexpected condition changes compared with environments during the design phase. Such environment changes are more critical to AI-enabled MSF systems due to the data-driven nature of ML and DL. For instance, an autonomous driving system's object detector might be trained with data only collected from sunny days. While the autonomous driving system is expected to be safe and reliable during rainy days, however, it is hard to determine to what extent the system can handle such a weather change. That is, the weather change in the open environment could result in corrupted sensor signals, leading to potential distribution changes of data that affect an MSF system's performance. To evaluate an MSF system's performance against such operational environments' changes, collecting and labeling real-world data is ideal but not feasible. To address these, we collect and design thirteen corruption patterns (Table~\ref{tab: corruption}) to synthesize corrupted signals for MSF systems, which can be grouped into three categories: (1) weather corruption, (2) sensor corruption, and (3) sensor misalignment. Weather corruptions represent the external environment changes of an MSF system, e.g., rainy/foggy days and bright/dark light conditions for a self-driving car, a UAV, etc. Sensor corruptions reflect the internal environment changes of an MSF system, such as transmission noise. Sensor misalignment is specifically designed for MSF systems given that the fusion of different signals requires accurate temporal and spatial calibration. Now we briefly introduce each category.

\subsubsection{Weather Corruption}
Weather conditions are an important factor that can inevitably affect the sensor's perception in the open environment, resulting in the performance degradation of MSF systems.
For example, normal cameras could hardly perceive the surroundings at night. In this work, we leverage weather corruption patterns from two perspectives: (1) light conditions change, and (2) adverse weather conditions. 

\begin{table}[t]
\caption{Corruption patterns used in this study.}
\label{tab: corruption}
\centering
\renewcommand{\arraystretch}{1.2}
\small
\begin{tabular}{llc}
\toprule
\textbf{Category}                                                              & \textbf{Corruption}             & \textbf{Modality} \\
\midrule
                                                                                                       & Rain (RN)                       & Camera \& LiDAR            \\
                                                                                                       & Fog (FG)                        & Camera \& LiDAR            \\
                                                                                                       & Brightness (BR)                 & Camera              \\
\multirow{-4}{*}{\begin{tabular}[c]{@{}l@{}}Weather\\ Corruption\end{tabular}}                         & Darkness (DK)                   & Camera              \\
\rowcolor{lightgray} 
\cellcolor{lightgray}                                                                               & Distortion (DT)                 & Camera              \\
\rowcolor{lightgray} 
\cellcolor{lightgray}                                                                               & Motion Blur (MB)                & Camera              \\
\rowcolor{lightgray} 
\cellcolor{lightgray}                                                                               & Defocus Blur (DB)               & Camera              \\
\rowcolor{lightgray} 
\cellcolor{lightgray}                                                                               & Image Gaussian Noise (GN)       & Camera              \\
\rowcolor{lightgray} 
\cellcolor{lightgray}                                                                               & Point Cloud Gaussian Noise (GN) & LiDAR              \\
\rowcolor{lightgray} 
\cellcolor{lightgray}                                                                               & Image Impulse Noise (IN)        & Camera              \\
\rowcolor{lightgray} 
\multirow{-7}{*}{\cellcolor{lightgray}\begin{tabular}[c]{@{}l@{}}Sensor \\ Corruption\end{tabular}} & Point Cloud Impulse Noise (IN)  & LiDAR              \\
                                                                                                       & Spatial Misalignment (SM)       &  Camera \& LiDAR             \\
\multirow{-2}{*}{\begin{tabular}[c]{@{}l@{}}Sensor \\ Misalignment\end{tabular}}                       & Temporal Misalignment (TM)      &          Camera \& LiDAR    
\\\bottomrule
\end{tabular}
\end{table}

\vspace{1mm}
\noindent\textbf{Lighting conditions}.
The camera is sensitive to lighting conditions, variations in daylight and road illumination can easily affect the image quality, while lighting conditions' effects on LiDAR are limited~\cite{survey_feng2020deep}. Therefore, we mainly focus on adjusting the \textbf{brightness (BR)} and \textbf{darkness (DK)} of the image pixels. 

\vspace{1mm}
\noindent\textbf{Weather conditions}.
Adverse weather can cause asymmetric measurement distortion of sensors, which poses a significant challenge for MSF perception systems that rely on redundant information.
For example, on rainy days, raindrops could lead to pixel attenuation and rain streaks on the image, meanwhile, the droplets will make the laser scattering and absorption, resulting in a lower intensity of points and perceived quality of LiDAR. 

\begin{figure*}[t]
\centering     
\subfigure[Fog]{\label{fig:a}\includegraphics[width=0.4\linewidth]{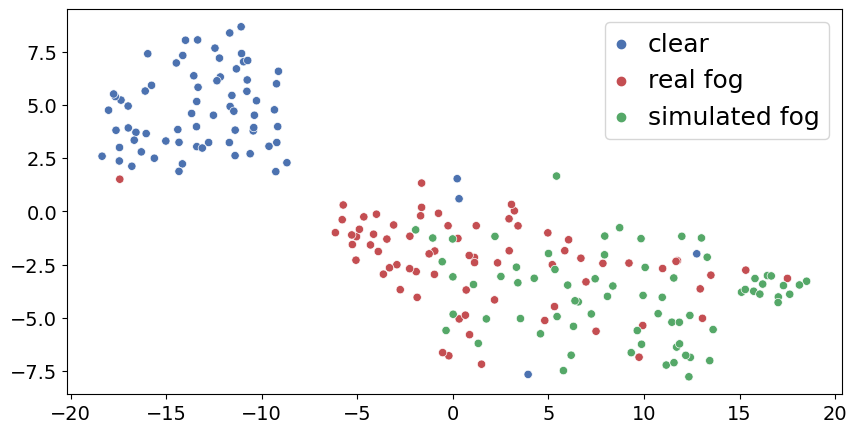}}
\hspace{10mm}
\subfigure[Rain]{\label{fig:b}\includegraphics[width=0.4\linewidth]{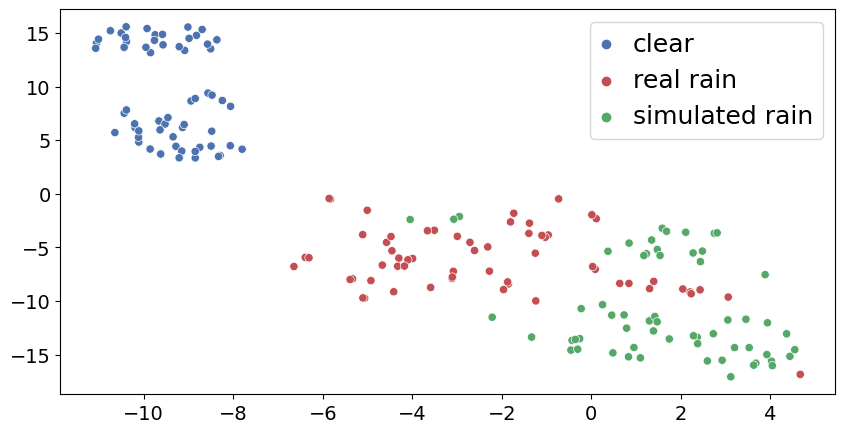}}
\vspace{-10pt}
\caption{Feature visualization of simulated rain/fog by T-SNEs.}
\label{fig: sim_rain_feature}
\end{figure*}

In our benchmark, we choose the domain-specific physical model to simulate the properties of two representative adverse weather, i.e., \textbf{rain (RN)} and \textbf{fog (FG)}. 
Specifically, we adopt rain model described in \cite{rain_halder2019physics} and fog model described in \cite{israel1959koschmieders} for camera, and rain/fog model described in \cite{rain_kilic2021lidar} for LiDAR.
Another critical problem when designing rain or fog corruptions is ensuring different sensors are sensing identical environments, e.g., the camera and LiDAR are both sensing a rain of 10mm/h. To address this, we control the environmental parameters in LiDAR and camera model to ensure the consistency of the rain's volumes or fog's maximum visibility.



\vspace{1mm}
\noindent\sloppy {\bf Realisticness validation of Rain and Fog corruption}. To validate the naturalness of rain and fog corruptions, we train deep fusion-based classifiers to distinguish real rain/fog scenes from clean scenes using datasets collected from real rainy/foggy weather~\cite{bijelic2020seeing, burnett2022boreas}. Then, we use these trained classifiers to make predictions on simulated data to measure the similarity between simulated and real data. We retrain each classifier five times and take the averaged accuracy.
The average classification accuracy of these trained weather classifiers is 98.6\% and 98.0\% on simulated fog/rain data. These results confirm that the simulated fog/rain data are highly similar compared to the real data.

We further analyze the similarity between the semantic features' distribution of real and simulated data. Specifically, we extract the high-level semantic features from the trained classifier. Then, we utilize T-SNE \cite{van2008visualizing} to reduce the dimensionality of acquired features to 2 and visualize these 2D features. As shown in Figure~\ref{fig: sim_rain_feature}, the distributions of the real and simulated corruptions are similar.



\subsubsection{Sensor Corruption}
Sensor corruptions reflect internal environment changes that lead to corrupted sensor signals, e.g., noises during transmission, and sensor artifacts that lead to blurry images. In this benchmark, we consider sensor corruption from two perspectives: (1) noise pattern, and (2) sensor artifacts.

\vspace{1mm}
\noindent\textbf{Noise Pattern}. Noise typically exists in both camera and LiDAR~\cite{duan2021low}.
There are two main sources of noise, one is from the sensor itself, such as sensor vibration~\cite{wang2021simultaneous}, random reflections and the low-ranging accuracy of LiDAR lasers~\cite{ma2012analysis}.
The other is due to the digital signal in its transmission recording process~\cite{exp_noise_wolff2016point}. 
We leverage two of the most common noise for each sensor, i.e., \textbf{Gaussian noise (GN)} and \textbf{impulse noise (IN)}.
Specifically, Gaussian noise applies Gaussian distributed noise to each point's coordinate in a point cloud or each pixel's value in an image. Impulse noise applies deterministic perturbations to a subset of points or randomly changes the value of image pixels.

\vspace{1mm}
\noindent\textbf{Sensor Artifacts}.
Sensor corruption could also result in artifacts of sensing results. For instance, \textbf{defocus blur (DB)} occurs when a camera is out of focus~\cite{benchmark_cifar10}; \textbf{motion blur (MB)} appears when a camera is shaking or moving quickly~\cite{benchmark_cifar10}. 
\textbf{Distortion (DT)} is one of the common basic optical aberrations caused by the optical design of lenses~\cite{zhao2021joint}. Note that, as an early attempt, we only consider artifacts of camera sensors. We leave artifacts of LiDAR sensors, e.g., one of LiDAR's beams is broken, as the future work.

\subsubsection{Sensor Misalignment}
Well-calibrated and synchronized sensors are a prerequisite for MSF-based perception systems. However, it is not easy to guarantee the perfect alignment of sensors in the real world~\cite{survey_feng2020deep}. 
Therefore, we design two corruption patterns, \textbf{Spatial misalignment (SM)} and \textbf{Temporal misalignment (TM)}, to simulate the misalignment between the camera and LiDAR.

\vspace{1mm}
\noindent\textbf{Spatial misalignment}.
MSF system requires an external calibration of each sensor during the assembly process to ensure that the position measured in different coordinate systems can be converted to each other.
However, even with well-calibrated sensors, the position of the sensors can inevitably deviate due to mechanical vibrations (e.g., when a self-driving car rides on a bumpy road) and thermal fluctuations~\cite{9196627}. Suppose a 3D point in the LiDAR coordinate is $\mathbf{p}_{li}$ and a corresponding point in the camera coordinate is $\mathbf{p}_{cam}$. The transformation from the LiDAR coordinate to the camera one can be expressed as: 
\begin{equation}
\small
\mathbf{p}_{cam}= \mathbf{T}_{\text {velo }}^{\text {cam }} \mathbf{p}_{li}
\end{equation}
where $\mathbf{T}_{\text {velo }}^{\text {cam }} $ is a rigid body transformation matrix.
In our experiments, we add a minor rotation (within $2^{\circ}$) to each rotation angle (i.e., roll, yaw, pitch) to simulate spatial misalignment between the camera and LiDAR.

\vspace{1mm}
\noindent\textbf{Temporal misalignment}.
MSF system requires synchronization of sensors to ensure the output from each individual branch is sensed at the same time.
In practical scenarios, sensor or transmission failure may cause a delay in one branch, resulting in a temporal misalignment~\cite{yeong2021sensor}. To simulate temporal misalignment, for a timestamp $t_o$, we replace the data $M_{i}(t_o)$ with the $M_{i}(t_o - \Delta t)$. This could represent a signal delay of $\Delta t$ second on branch $i$.

\subsection{Evaluation Metrics}
\label{sec: eval metrics}
Our benchmarks provide specific quantitative performance evaluation metrics for each perception task, including object detection, object tracking, and depth completion. Then, we define robustness evaluation metrics based on these metrics.
Below we describe each perception task and the corresponding evaluation metrics. 
 
\textbf{Object detection} aims to locate, classify and estimate oriented bounding boxes in the 3D space. Note that in this benchmark, we mainly evaluate the detection of \textit{Car} objects with moderate difficulty. The accuracy of object detection can be measured by IOU (intersection over union) and AP (average precision).

IOU measures the overlap area between a ground-truth 3D bounding box $B_{g}$ and a predicted 3D bounding box $B_{p}$ over their union\cite{iou_padilla2020survey}. The computation of IOU can be represented as:
\begin{equation}
\small
IOU=\frac{\operatorname{area}\left(B_{p} \cap B_{g}\right)}{\operatorname{area}\left(B_{p} \cup B_{g}\right)}
\end{equation}
In our experiments evaluation, we define a successful detection as an IOU larger than 70\%.

AP is used to measure the performance of the overall detection performance, which approximates the shape of the Precision/Recall curve as:
\begin{equation}
\small
\left.\mathrm{AP}\right|_{R}=\frac{1}{|R|} \sum_{r \in R} \rho_{\text {interp }}(r)
\end{equation}
\sloppy We apply forty equally spaced recall levels~\cite{ap_simonelli2019disentangling}, i.e., $R_{40}=$ $\{1 / 40,\;2 / 40,\;\ldots,\;1\}$. The interpolation function is defined as: 
$\rho_{\text {interp }}(r)=\max _{r^{\prime}: r^{\prime} \geq r} \geq\left(r^{\prime}\right)$, 
where $\rho(r)$ gives the precision at $r$.

\textbf{Multiple object tracking} aims to maintain objects' identities and track their location across data frames over time. The accuracy is measured by MOTA (multiple object tracking accuracy)~\cite{mota_bernardin2008evaluating}:
\begin{equation}
\small
    \mathrm{MOTA}=1-\frac{\sum_{t}\left(\mathrm{FN}_{t}+\mathrm{FP}_{t}+\mathrm{IDSW}_{t}\right)}{\sum_{t} \mathrm{GT}_{t}}
\end{equation}
where $\mathrm{{FN}_{t}}$, $\mathrm{{FP}_{t}}$, and $\mathrm{IDSW}_{t}$ are the number of misses, of false positives, and of mismatches, respectively, during a period $t$. The MOTA can be regarded as a measurement of three different types of errors. 
 
\textbf{Depth completion} aims to up-sample sparse irregular depth to dense regular depth. The depth completion tasks focus on predicting the distance for every pixel in the image from the viewer given LiDAR point cloud and image data. We use the Root Mean Squared Error (RMSE, mm) to measure the distance between the predicted depth and ground-truth value:
\begin{equation}
\small
    R M S E=\sqrt{\frac{1}{m} \sum_{i=1}^{m}\left(d^{i}_{p} - d^{i}_{g}\right)^{2}}
\end{equation}
where $d^{i}_{p},d^{i}_{g}$ are the predicted depth and ground-truth of the $i$th position, $m$ is the total number of ground-truth.

 
To further evaluate the robustness of different fusion mechanisms across different MSF systems and tasks, we define the robustness of MSF on a corruption pattern $c \in C$ with severity $s \in S$ as its performance $P_{c}^{s}$ relative to $P_{clean}$ (performance on clean data): 
\begin{equation}
\small
Rb_{c}^{s} = P_{c}^{s}/P_{clean}
\end{equation}
where $P$ is measured by one of the evaluation metrics for the corresponding MSF task, i.e., AP, MOTA, or RMSE~\footnote{Note that we normalize the metric of each task into $[0, 1]$.} and $clean$ represents the clean data. 
A larger $Rb_{c}$ means that the system's performance against a specific corruption pattern is closer to the normal performance.
Then, we estimate the robustness of an MSF system by averaging over all of the corruption patterns $c$ with severity $s$, i.e.
\begin{equation}
\small
    mRb =\frac{1}{|S|} \sum_{s \in S} \frac{1}{|C|} \sum_{c \in C} Rb_{c}^{s}
\end{equation}
A lower $mRb$ means a higher risk of performance degradation when the MSF system is deployed in the open operational environment.

Note that both $Rb_{c}$ and $mRb$ can generalize to different MSF systems, tasks, and corruption patterns. In this way, we expect our benchmark and evaluation metrics to be flexible and extensible.

\subsection{Dataset}
\label{sec: dataset}
KITTI~\cite{kitti_geiger2012we} is one of the most popular autonomous driving datasets, which adopts four high-resolution cameras, a Velodyne HDL-64E LiDAR, and an advanced positioning system to collect data from different real-world driving scenarios.
KITTI supports diverse perception tasks, including 3D object detection, 3D object tracking, depth completion, etc.
During the paper collection process, we also found that more than two-thirds of the MSF perception systems are evaluated on KITTI. 
To this end, we use the KITTI as our base dataset to construct KITTI-C to benchmark AI-enabled MSF systems' performance and robustness. Note that, corruption patterns used in this study can also generalize to other datasets, such as Waymo~\cite{waymo_sun2020scalability} and NuScenes~\cite{nuScenes_caesar2020nuscenes}.




\section{Empirical Study Design}

In this section, we introduce our research questions and experimental setup. We first investigate the robustness of existing AI-enabled MSF systems from three perspectives: (1) against corrupted signals (\textbf{RQ1}), (2) against spatial/temporal misalignments (\textbf{RQ2}), and (3) against partial/complete signal loss (\textbf{RQ3}). Then, we investigate the potential of repairing these MSF systems' robustness (\textbf{RQ4}).


\begin{figure*}[t]
  \centering 
  \subfigure[EPNet]{
  \includegraphics[width=0.2\linewidth]{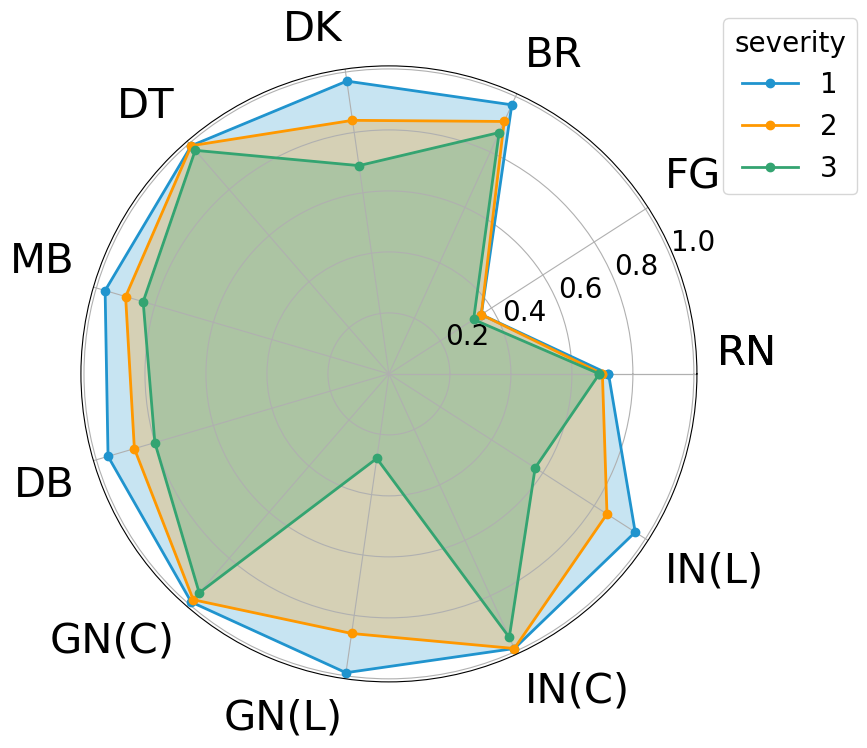}
  }
  \subfigure[CLOCs]{
  \includegraphics[width=0.2\linewidth]{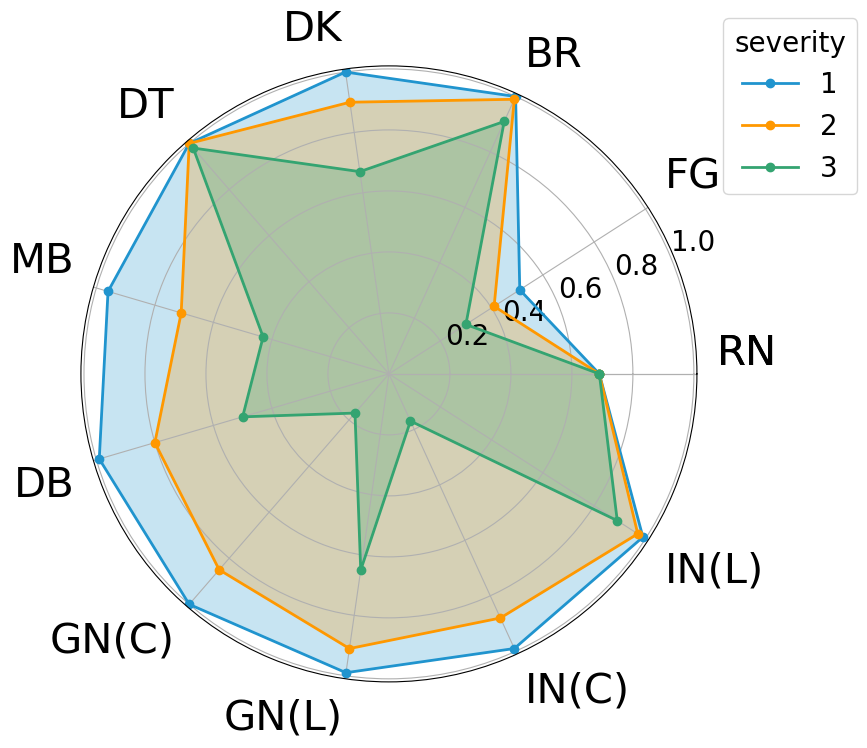}
  }
  \subfigure[FConv]{
  \includegraphics[width=0.2\linewidth]{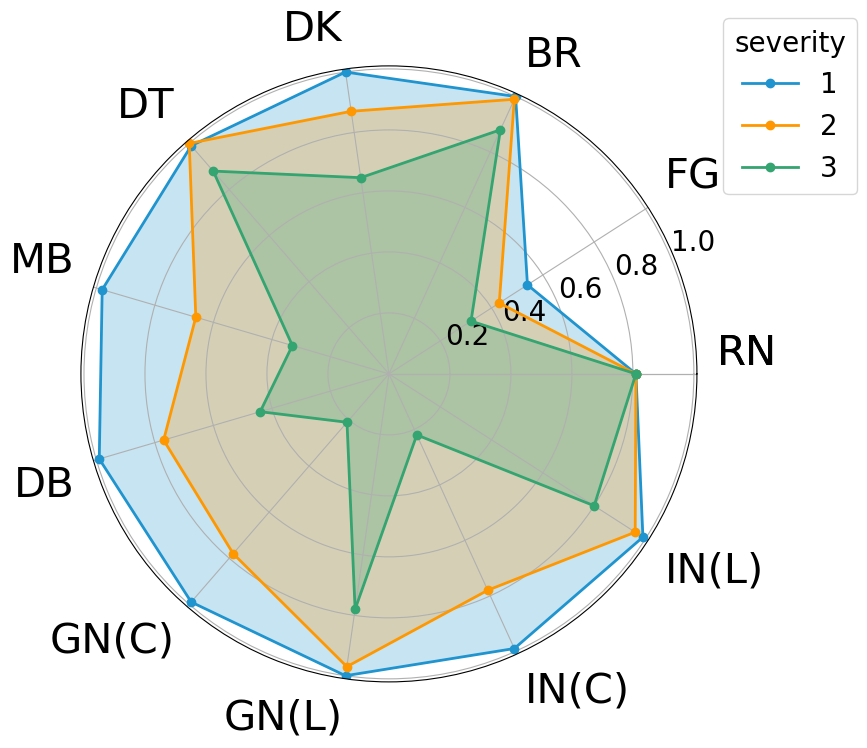}
  }
  \subfigure[JMODT]{
  \includegraphics[width=0.2\linewidth]{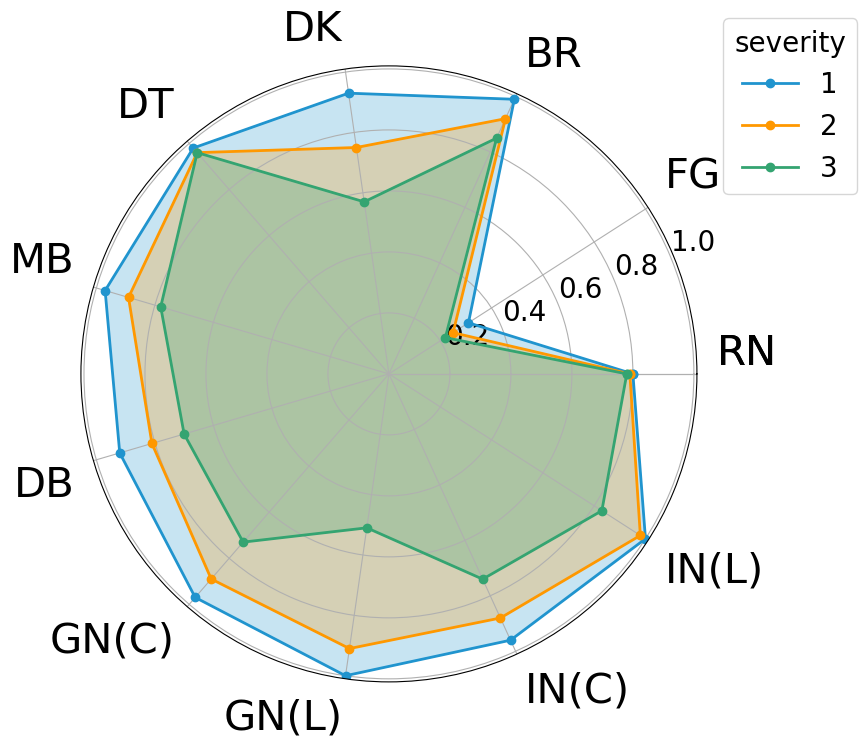}
  }
  \subfigure[DFMOT]{
  \includegraphics[width=0.2\linewidth]{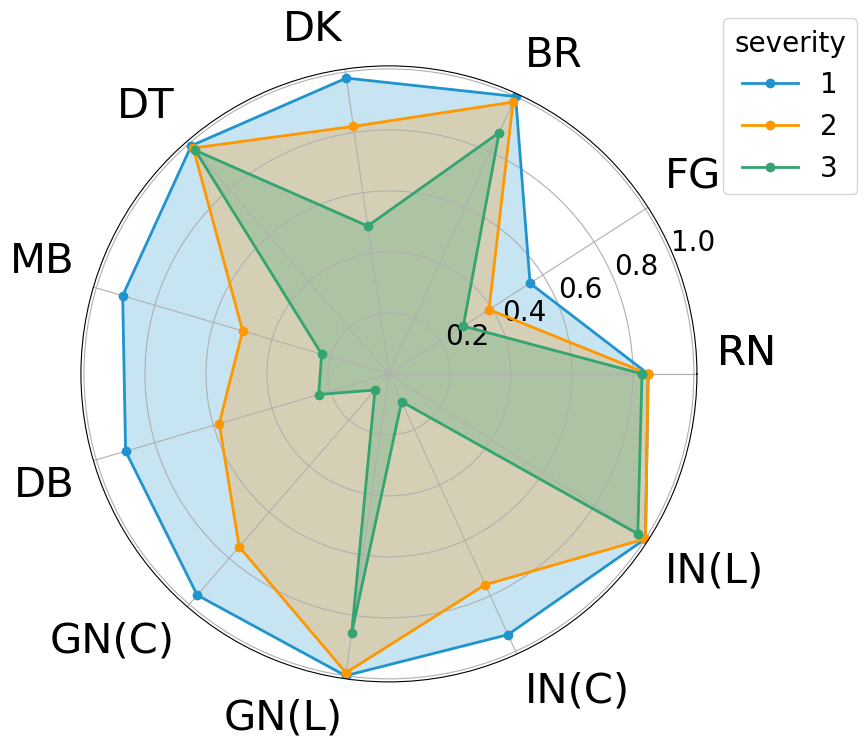}
  }
  \hspace{5mm}
  \subfigure[TWISE]{
  \includegraphics[width=0.2\linewidth]{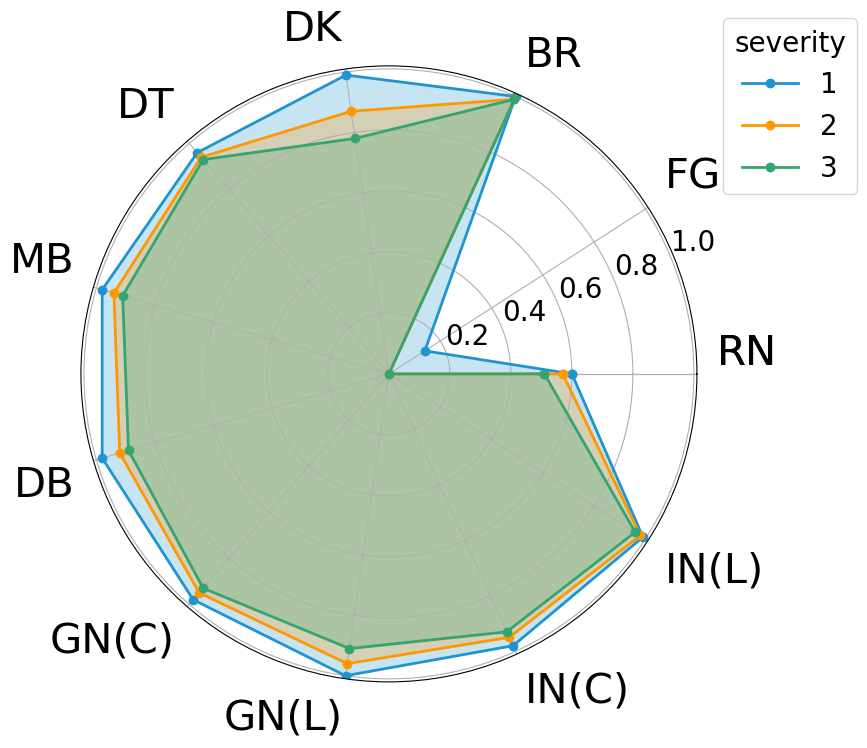}
  }
  \hspace{5mm}
  \subfigure[MDANet]{
  \includegraphics[width=0.2\linewidth]{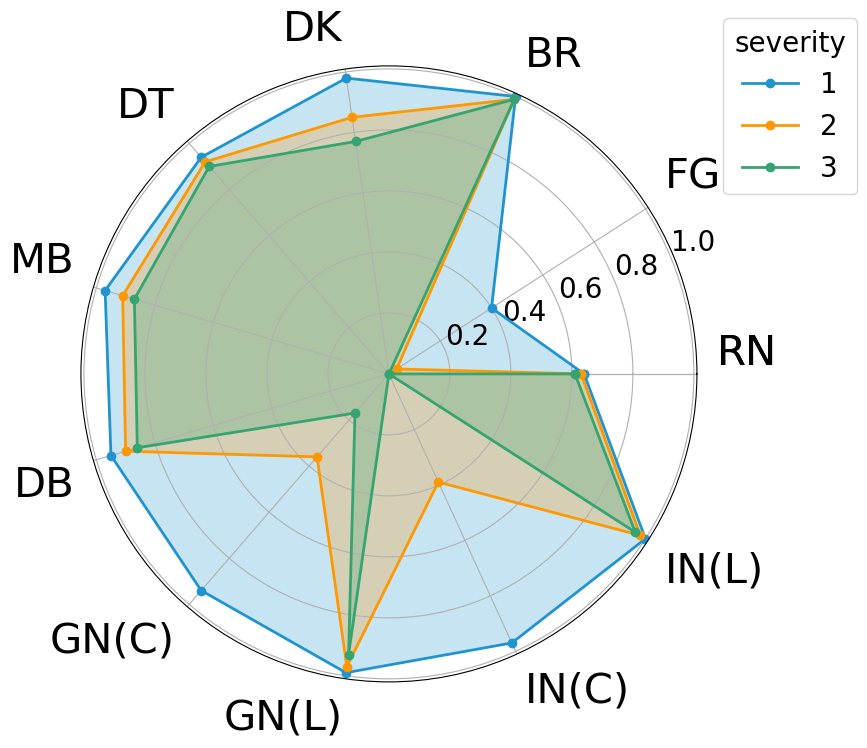}
  }
  \caption{Robustness performance of seven MSF systems against different corruption patterns.}
  \label{fig: RQ MSF rob}
\end{figure*}

\begin{table*}[t]
\small
\renewcommand{\arraystretch}{1.2}
\caption{Average robustness performance of MSF systems against different corruption patterns across three severity levels.}
\label{tab: rq1_rob}
\begin{tabular}{cc|cccc|ccc|cccc|cccc}
\toprule
\multicolumn{2}{c|}{\multirow{2}{*}{\textbf{Task}}}   & \multicolumn{4}{c|}{\textbf{\begin{tabular}[c]{@{}c@{}}Weather\end{tabular}}} & \multicolumn{3}{c|}{\textbf{\begin{tabular}[c]{@{}c@{}}Sensor \end{tabular}}} & \multicolumn{4}{c|}{\textbf{\begin{tabular}[c]{@{}c@{}}Noise\end{tabular}}} & \multirow{2}{*}{\textbf{$Rb^{s1}$}} & \multirow{2}{*}{\textbf{$Rb^{s2}$}} & \multirow{2}{*}{\textbf{$Rb^{s3}$}} & \multirow{2}{*}{\textbf{mRb}} \\

\multicolumn{2}{c|}{}                                 & RN                   & FG                   & BR                   & DK                   & DT                           & MB                           & DB                          & GN(C)                & GN(L)                & IN(C)               & IN(L)               &                               &                               &                               &                              \\
\midrule
\multirow{3}{*}{\textbf{Object}}   & \textbf{EPNet}  & 0.71                 & 0.35                 & 0.92                 & 0.83                 & 0.98                         & 0.90                         & 0.88                        & 0.98                 & 0.71                 & 0.98                & 0.79                & 0.90                          & 0.84                          & 0.72                          & 0.82                         \\
                                   & \textbf{FConv}  & 0.81                 & 0.43                 & 0.96                 & 0.84                 & 0.96                         & 0.66                         & 0.73                        & 0.66                 & 0.92                 & 0.66                & 0.92                & 0.93                          & 0.82                          & 0.58                          & 0.78                         \\
                                   & \textbf{CLOCs}  & 0.69                 & 0.41                 & 0.97                 & 0.85                 & 0.99                         & 0.70                         & 0.76                        & 0.67                 & 0.85                 & 0.68                & 0.95                & 0.92                          & 0.83                          & 0.58                          & 0.77                         \\

\midrule
\multirow{2}{*}{\textbf{Tracking}} & \textbf{JMODT}  & 0.79                 & 0.26                 & 0.92                 & 0.75                 & 0.97                         & 0.88                         & 0.81                        & 0.86                 & 0.81                 & 0.86                & 0.94                & 0.89                          & 0.82                          & 0.70                          & 0.80                         \\
                                   & \textbf{DFMOT}  & 0.84                 & 0.41                 & 0.95                 & 0.77                 & 0.98                         & 0.54                         & 0.57                        & 0.59                 & 0.95                 & 0.60                & 0.99                & 0.92                          & 0.78                          & 0.54                          & 0.75                         \\
\midrule
\multirow{2}{*}{\textbf{Depth}}    & \textbf{TWISE}  & 0.56                 & 0.05                 & 0.99                 & 0.88                 & 0.94                         & 0.94                         & 0.93                        & 0.95                 & 0.95                 & 0.95                & 0.98                & 0.87                          & 0.83                          & 0.79                          & 0.83                         \\
                                   & \textbf{MDANet} & 0.62                 & 0.14                 & 0.99                 & 0.86                 & 0.92                         & 0.92                         & 0.90                        & 0.49                 & 0.96                 & 0.46                & 0.98                & 0.89                          & 0.72                          & 0.64                          & 0.75                         \\
\midrule
\multicolumn{2}{c|}{\textbf{Avg}}                     & 0.72                 & 0.29                 & 0.96                 & 0.83                 & 0.96                         & 0.79                         & 0.80                        & 0.74                 & 0.88                 & 0.74                & 0.94                & 0.90                          & 0.81                          & 0.65                          & 0.67                        
\\\bottomrule
\end{tabular}
\end{table*}

\subsection{Research Questions}

{\noindent\rqOne} Though a few AI-enabled MSF perception systems have been proposed and used, there is no systematic study on the robustness
of these systems. 
In this RQ, we focus on corrupted signals due to weather, sensor, and noise corruptions (Table~\ref{tab: corruption}). For each corruption pattern, we adopt three different levels of severity. Specifically, for rain and fog, three severity levels represent 10mm/h, 25mm/h, and 50mm/h of rainfall and 104m, 80m, and 51m of visibility, respectively. 
To sum up, we conduct experiments with 231 different configurations (11 corruptions $\times$ 3 levels $\times$ 7 MSF systems) to investigate this RQ. 

\vspace{1mm}
{\noindent \rqTwo}
RQ2 aims to evaluate the AI-enabled MSF system's sensitivity to calibration errors.
To simulate the spatial misalignment, 
we rotate the LiDAR sensor around the x, y, and z axes by $0.5^\circ$, $1^\circ$, and $2^\circ$, respectively.
To simulate temporal misalignment, we create five levels of LiDAR and camera signal delay, 
i.e., 0.1s, 0.2s, \ldots, 0.5s, 
respectively.
We only investigate temporal misalignment's effects on object tracking systems as the other two tasks are not time-sensitive.

\vspace{1mm}
{\noindent\rqThree}
This RQ aims to investigate how existing AI-enabled MSF systems are coupled and if they are robust enough against signal loss of one source of signals. To investigate this RQ, we simulate the signal loss with five different levels (10\%, 25\%, 50\%, 75\%, 100\%) of each branch. For the camera branch, we reshape the image into a one-dimensional array and randomly drop pixels. For the LiDAR branch, we randomly remove points with different percentages.

\vspace{1mm}
{\noindent\rqFour}
RQ4 aims to investigate the properties of different fusion mechanisms, and analyze the weakness or potential threats of each based on the experiment results from RQ1-3.
We first divide MSF systems into three categories according to their fusion mechanisms. To further investigate the possibility of repairing MSF systems, we make an early attempt on enhancing MSF systems' robustness by improving the fusion mechanism of late and weak fusion.

\subsection{Experimental Setup}



In experiments, we use 
Second~\cite{yan2018second} as the LiDAR branch for CLOCs and DFMOT, Cascade-RCNN~\cite{cai2018cascade} as the camera branch for CLOCs, DFMOT, and FConv. 
We implement all MSF systems with PyTorch 1.8 and Python 3.7. For each system, we use default configurations to ensure a consistent runtime environment.
Table~\ref{tab: fusion systems} shows the performance of each reproduced system.
The detailed settings of each system can be found in supplementary website~\cite{website}.
All experiments are conducted on a server with an Intel i7-10700K CPU (3.80 GHz), 48 GB RAM, and an NVIDIA RTX 3070 GPU (8 GB VRAM).

\section{Experimental Results}
\subsection{RQ1. AI-enabled MSF is not robust against corrupted signals.}

Fig.~\ref{fig: RQ MSF rob} summarizes the robustness benchmark results for seven AI-enabled MSF perception systems against eleven corruption patterns via radar charts. Each axis in the figure represents the robustness score $Rb^{s}_{c}$ against corruption $c$ with severity level $s$. These results reveal that all the selected AI-enabled MSF systems have robustness issues against corrupted signals, while their robustness properties could be varied. For instance, all the selected systems perform poorly against fog (FG) corruption. However, for the blur effects (MB, DB), some systems perform relatively robust (e.g., EPNet, TWISE, JMODT), while some face severe robustness issues (e.g., CLOCs, FConv, DFMOT). To further analyze how different MSF systems perform against different categories of corrupted signals, we interpret the detailed robustness performance in Table~\ref{tab: rq1_rob} by presenting the average performance against each corruption pattern across three severity levels. 

\textbf{Weather Corruption}. As shown in Table~\ref{tab: rq1_rob}, weather corruptions pose significant robustness issues for MSF systems, where the average robustness score against rain (RN) and fog (FG) are 0.72 and 0.29, respectively. We also find that the depth completion systems (i.e., TWISE, MDANet) hardly work on foggy days. Specifically, the highest robustness score among depth completion systems is only 0.14. Besides, decreasing brightness affects MSF systems more significantly compared with increasing brightness, where the average robustness scores are 0.83 and 0.96, respectively.

\textbf{Sensor Artifact}. While all the MSF systems are relatively robust against distortion (robustness score higher than 0.9), some of them (i.e., FConv, CLOCs, DFMOT) particularly have significant performance degradation against blur effects (MB, DB). We further qualitatively check the image signals corrupted by distortion (DT) and find that only the edges of images are distorted. This could be one possible reason that the effects of DT are limited.

\textbf{Noise Corruption}.
As shown in Table \ref{tab: rq1_rob}, camera signals corrupted by noise patterns are usually more vulnerable in MSF systems, where the robustness score against noises in cameras (74.4 (GN), 74.2 (IN)) are lower than those in LiDAR (87.9 (GN), 89.6 (IN)). Based on these observations, adding appropriate filters for image signals could be important for designing robust AI-enabled MSF.

\begin{center}
\vspace{1mm}
\fcolorbox{black}{gray!10}{\parbox{.95\linewidth}{\textbf{Answer to RQ1:}
    Existing AI-enabled MSF systems are not robust enough against common corruption patterns. 
   Moreover, among the 11 common corruptions, adverse weather causes the most severe robustness degradation.
  }}
\end{center}

\subsection{RQ2. AI-enabled MSF is sensitive to sensor misalignment.}

Through our investigation of RQ2, we find that AI-enabled MSF systems are sensitive to both spatial and temporal misalignment. 

\textbf{Spatial misalignment}. Table~\ref{tab: rq3_sp_mis} shows the experimental results of spatial misalignment, where each cell represents the robustness score. According to the average robustness score across different rotation axes and angles (last row of Table~\ref{tab: rq3_sp_mis}), we can find that spatial misalignment significantly affects MSF's robustness. Specifically, the highest average robustness score among the seven systems is lower than 0.78. We also find that MSF systems of different tasks could have different sensitivity regarding spatial misalignment. For instance, the robustness scores of object detection systems (i.e., EPNet, FConv, CLOCs) are relatively lower than object tracking and depth completion systems.

\begin{table}[t]
\renewcommand{\arraystretch}{1.2}
\small
 \caption{Robustness performance of MSF systems against spatial misalignment.}
 \label{tab: rq3_sp_mis}
\setlength{\tabcolsep}{2.2pt}{
\begin{tabular}{cc|ccccccc}
\toprule
\multicolumn{2}{c|}{\textbf{Axis}} & \textbf{EPNet} & \textbf{FConv} & \textbf{CLOCs} & \textbf{JMODT} & \textbf{DFMOT} & \textbf{TWISE} & \textbf{MDANet} \\
\midrule
\multirow{3}{*}{X}                           & 0.5°                           & 0.93           & 0.92           & 0.96           & 0.98           & 0.99           & 0.94           & 0.94            \\
                                             & 1°                             & 0.80           & 0.79           & 0.83           & 0.94           & 0.99           & 0.82           & 0.83            \\
                                             & 2°                             & 0.46           & 0.48           & 0.57           & 0.80           & 0.72           & 0.57           & 0.58            \\
\midrule
\multirow{3}{*}{Y}                   & 0.5°                                   & 0.45                 & 0.41                 & 0.41                    & 0.93                 & 0.84                 & 0.73                 & 0.77                 \\
                                     & 1°                                     & 0.09                 & 0.04                 & 0.04                    & 0.54                 & 0.75                 & 0.16                 & 0.34                 \\
                                     & \cellcolor{tab_red}2°                  & \cellcolor{tab_red}0 & \cellcolor{tab_red}0 & \cellcolor{tab_red}0.06 & \cellcolor{tab_red}0 & \cellcolor{tab_red}0 & \cellcolor{tab_red}0 & \cellcolor{tab_red}0 \\
\midrule                                       
\multirow{3}{*}{Z}                      & 0.5°                                & 0.93           & 0.92           & 0.96             & 0.98           & 0.99           & 0.94           & 0.94            \\
                                        & 1°                                  & 0.80           & 0.79           & 0.83             & 0.95           & 0.99           & 0.82           & 0.83            \\
                                        & 2°                                  & 0.44           & 0.48           & 0.56             & 0.80           & 0.72           & 0.57           & 0.59            \\
\midrule
\multicolumn{2}{c|}{Avg}                                                       & 0.54                 & 0.54                 & 0.58                    & 0.77                 & 0.77                 & 0.62                 & 0.65

\\\bottomrule
\end{tabular}
}
\end{table}

\begin{figure}[t]
  \centering
  \subfigure[Clean]{
  \includegraphics[width=0.2\linewidth]{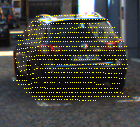}
  }
  \hspace{1mm}
  \subfigure[X-axis]{
  \includegraphics[width=0.2\linewidth]{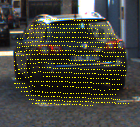}
  }
 \hspace{1mm}
  \subfigure[Y-axis]{
  \includegraphics[width=0.2\linewidth]{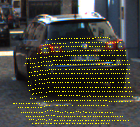}
  }
   \hspace{1mm}
  \subfigure[Z-axis]{
  \includegraphics[width=0.2\linewidth]{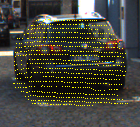}
  }

  \caption{An example of a $2^\circ$ rotation error in calibration around X-, Y-, and Z-axes.}
  \label{fig: RQ calib error}
\end{figure}

In addition, we also find that MSF systems are more sensitive to rotation around Y-axis. When the rotation angle around Y-axis is increased to $2^\circ$ (highlighted in Table~\ref{tab: rq3_sp_mis}), five out of seven systems crash (robustness score is 0), while the other two also have poor performance. By contrast, there is no such dramatic decrease for rotations around X- and Z-axes. We qualitatively compare the effects of $2^\circ$ rotation around different axes by projecting the point cloud onto the image in Fig.~\ref{fig: RQ calib error}. A $2^\circ$ rotation around Y-axis results in a significant malposition between the image and point cloud, which possibly leads to the system crash.

\begin{figure}[t]
    \centering
    \includegraphics[width = 0.9\linewidth]{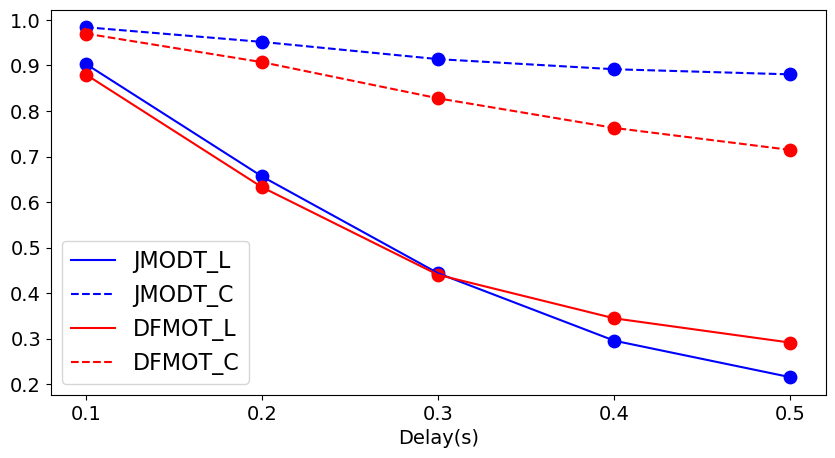}
    \caption{Robustness performance of MSF systems against temporal misalignment.}
    \label{fig: rq3_delay}
\end{figure}

\begin{figure*}[htbp]
  \centering
  \subfigure[Object Detection]{
  \includegraphics[width=0.3\linewidth]{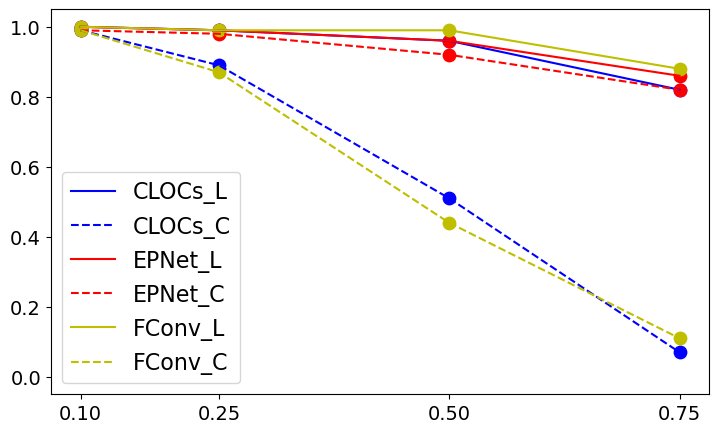}
  }
  \hspace{1mm}
  \subfigure[Object Tracking]{
  \includegraphics[width=0.3\linewidth]{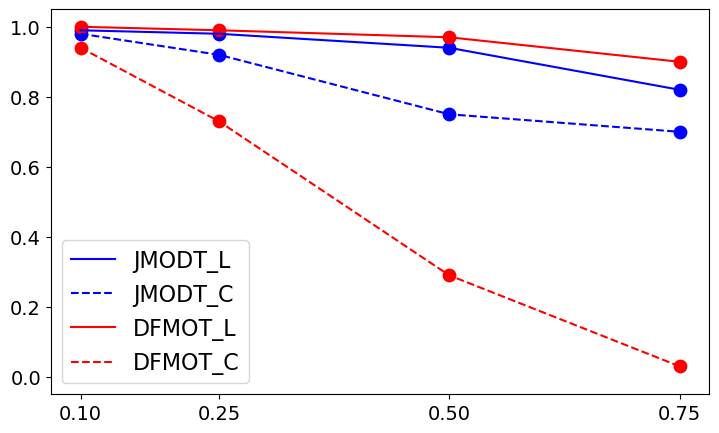}
  }
  \hspace{1mm}
  \subfigure[Depth Completion]{
  \includegraphics[width=0.3\linewidth]{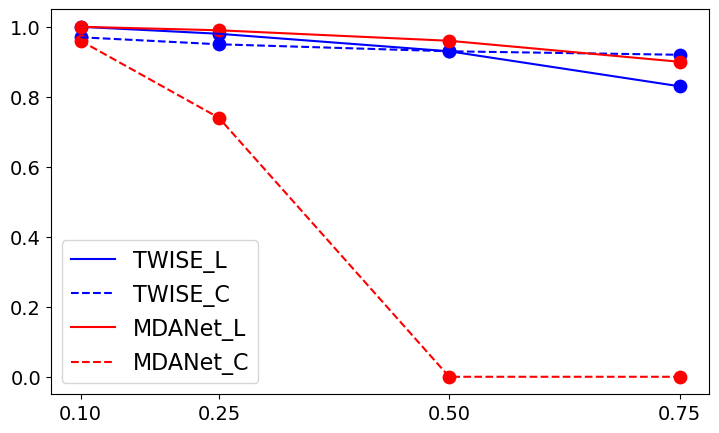}
  }

  \caption{
  MSF systems' performance when partially losing one source of the signals.
  }
  
  \label{fig: RQ signal loss}
\end{figure*}

\textbf{Temporal misalignment}. Fig.~\ref{fig: rq3_delay} shows the effects of temporal misalignment on AI-enabled MSF systems for object tracking (i.e., JMODT, DFMOT). As we can observe from Fig.~\ref{fig: rq3_delay}, both the camera and LiDAR branch are sensitive to the delay. When the delay increases, the robustness score of the MSF system decreases. In particular, we find that LiDAR is more sensitive (solid lines in Fig.~\ref{fig: rq3_delay}) to the delay. When the delay of LiDAR increases to 0.3 seconds, the robustness score of JMODT and DFMOT drops nearly 60\% (from 1.0 to 0.4). In contrast, the same level delay of the camera only drops their robustness performance by 10\%$\sim$20\%.

\begin{center}
\vspace{1mm}
\fcolorbox{black}{gray!10}{\parbox{.95\linewidth}{\textbf{Answer to RQ2:}
AI-enabled MSF perception systems are sensitive to both temporal and spatial misalignment, especially for LiDAR.
Even small synchronization (0.3 seconds) and calibration errors (2$^\circ$) can lead to a crash of AI-enabled MSF systems. 
  }}
\end{center}

\subsection{RQ3. Tightly-coupled AI-enabled MSF could be less robust.}
\label{sec: RQ2}

When deploying an AI-enabled MSF system, developers might expect it to be reliable even if one of the signals is lost. However, our experiments demonstrate that AI-enabled MSF systems are less robust as they crash when they partially or completely lose a source of signals. Fig.~\ref{fig: RQ signal loss} shows the robustness of different MSF systems with different severity levels of signal loss. These results suggest that partially losing either camera or LiDAR signal could affect the MSF system's performance, while losing the camera signal could be more critical (dashed-line in Fig.~\ref{fig: RQ signal loss}). Specifically, we find that losing the camera signal significantly affects 6 out of 7 systems (except EPNet) compared with losing the LiDAR signal. When losing 75\% of the camera signal, 4 out of 7 selected systems have a low robustness performance ($mRb$ smaller than 0.2). These results also suggest that existing MSF systems heavily depend on camera signals.



  

To further investigate AI-enabled MSF systems' robustness against signal loss, Table~\ref{tab: rq2_loss} shows the robustness performance of these systems when completely losing one source of the signal. We can find that when losing LiDAR signals, all of the systems crash. When losing camera signals, 3 out of 7 systems also crash and 2 systems have poor performance (e.g., EPNet, JMODT). Surprisingly, we find that MDANet does not crash when completely losing the camera signal, however, it crashes when losing partial signals (see Fig.~\ref{fig: RQ signal loss}c). One possible explanation is that due to the sparsity of objects in the image data, discarding 50\% or 75\% pixels could have dropped all the valuable information (e.g., pixels including objects). The remaining pixels, instead, could bring interference to the MSF system and thus lead to the system crash.

\begin{center}
\vspace{1mm}
\fcolorbox{black}{gray!10}{\parbox{.95\linewidth}
{\textbf{Answer to RQ3:}
AI-enabled MSF systems could be vulnerable when partially or completely losing one source of signals, even if the other source is working properly.
In particular, partially losing camera signals could be more critical for AI-enabled MSF systems. 
We also find that though tightly-coupled AI-enabled MSF systems have promising performance, they could be less robust when completely losing either camera or LiDAR signals.
  }}
\end{center}

\begin{table}[tb]
\footnotesize
\renewcommand{\arraystretch}{1.2}
\caption{MSF systems' performance when completely losing one source of the signals.}
\label{tab: rq2_loss}
\setlength{\tabcolsep}{3.5pt}{
\begin{tabular}{c|ccccccc}
\toprule
\textbf{Modality} & \multicolumn{1}{l}{{EPNet}} & \multicolumn{1}{l}{{FConv}} & \multicolumn{1}{l}{{CLOCs}} & \multicolumn{1}{l}{{JMODT}} & \multicolumn{1}{l}{{DFMOT}} & \multicolumn{1}{l}{{TWISE}} & \multicolumn{1}{l}{{MDANet}} \\
\midrule
C                 & 0.23                               & 0                                  & 0                                  & 0.13                               & 0                                  & 0.50                               & 0.58                                \\
L                 & 0                                  & 0                                  & 0                                  & 0                                  & 0                                  & 0                                  & 0   
\\\bottomrule
\end{tabular}
}
\end{table}

\subsection{RQ4. Fusion mechanisms could affect AI-enabled MSF's robustness and reliability.}
\label{sec:rq4}


While there is no systematic evidence indicating that one specific fusion mechanism is the most robust and reliable, we particularly find that different fusion mechanisms may have unique advantages and potential threats due to their inherent properties. According to our findings from RQ1, three deep fusion MSF systems (i.e., EPNet, JMODT, TWISE) are more robust against blur images (MB, DB) and noise patterns (IN(C), IN(L)) than others (Table~\ref{tab: rq1_rob}). According to our finding from RQ3, these systems also perform robustly when partially losing camera signals (Fig.~\ref{fig: RQ signal loss}). Two late fusion MSF systems (i.e., ClOCs, DFMOT) show similar trends against corrupted signal (RQ1) and signal loss (RQ3). To further investigate the effect of the fusion mechanism on the robustness, we try to repair the badly performed late- and weak-fusion MSF system based on the inherent properties of different fusion mechanisms. 

\begin{figure}[t]
    \centering
    \includegraphics[width = 0.9\linewidth]{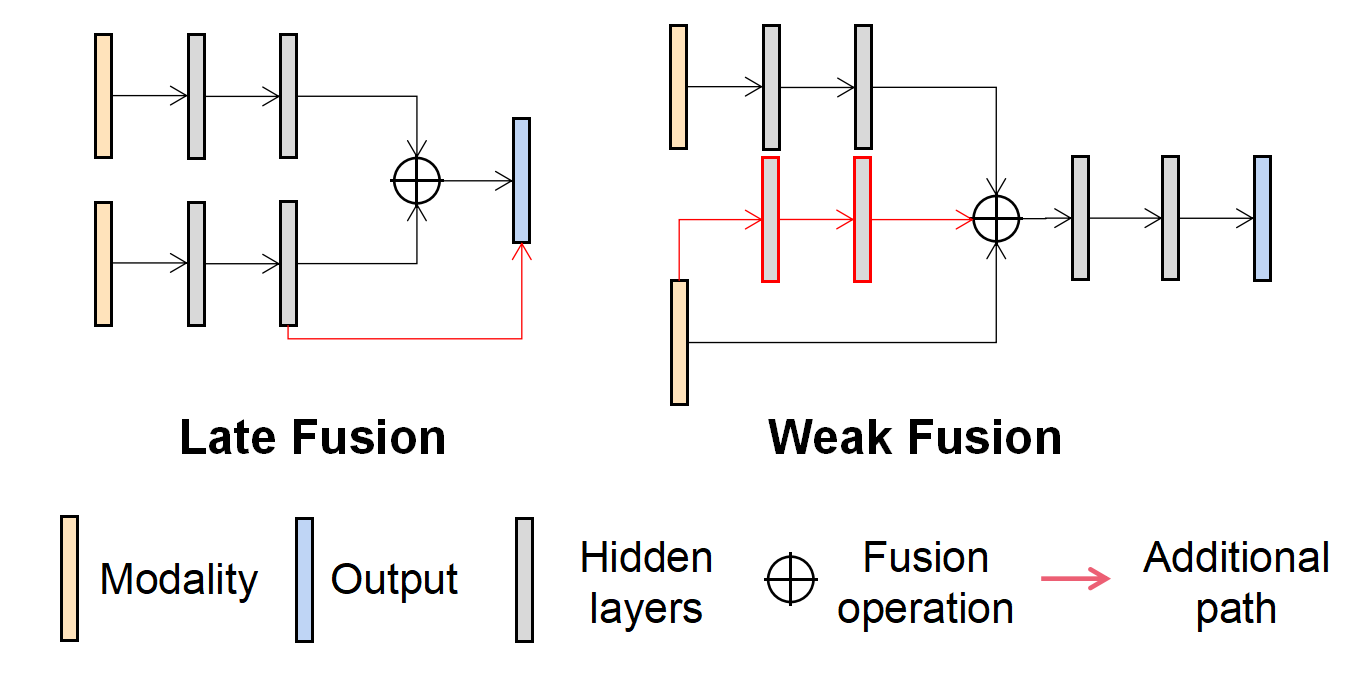}
    \caption{Improved AI-enabled MSF mechanisms.}
    \label{fig: fusion2}
\end{figure}
\vspace{-1mm}

To improve the late fusion, we leverage a shortcut between the LiDAR branch and the fusion layer to enhance the MSF robustness (left part of Fig.~\ref{fig: fusion2}).
Specifically, we design a matching method to aggregate high confidence and unique results from an individual branch to the fusion results.
This is motivated by our findings in RQ1 and RQ3, where the camera is more susceptible to external environmental interference. 

Weak fusion uses a cascade architecture to connect two modules in series. Its robustness performance bottleneck is due to inaccurate/missing guidance signals. Therefore, for weak fusion, we leverage a neural network to extract extra guidance from another modality and connect it to the downstream module as an additional guidance branch (right part of Fig.~\ref{fig: fusion2}). Specifically, we first train a 2D detector by projecting the point cloud to 2D front view images. Then we use the detecting results from the 2D front view as an extra guidance input.

\begin{figure}[t]
  \centering
  \subfigure[CLOCs]{
  \includegraphics[width=0.43\linewidth]{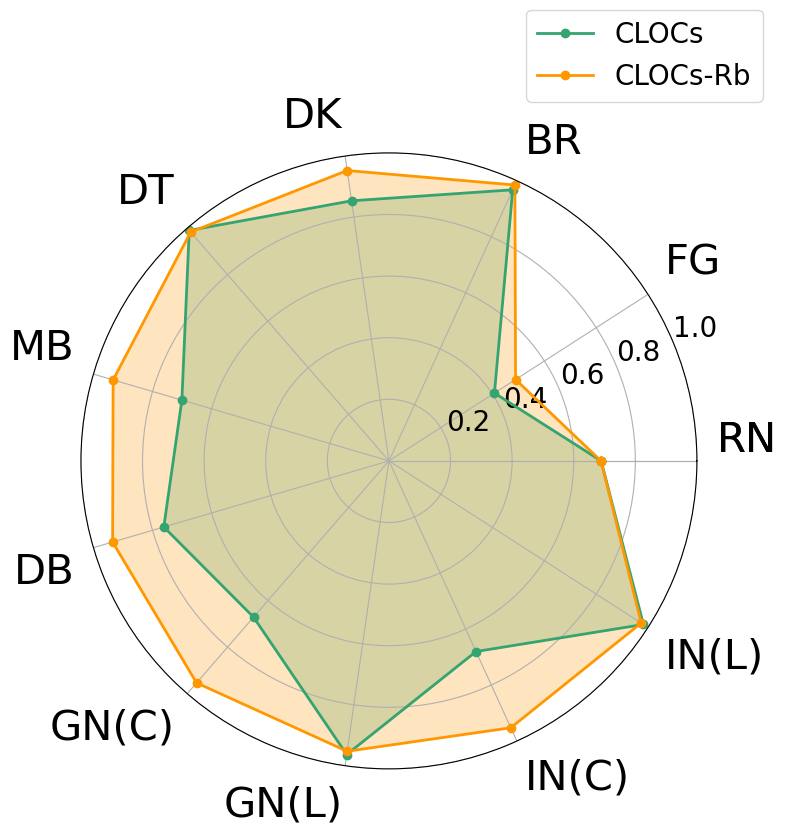}
  }
  \hspace{2mm}
  \subfigure[FConv]{
  \includegraphics[width=0.43\linewidth]{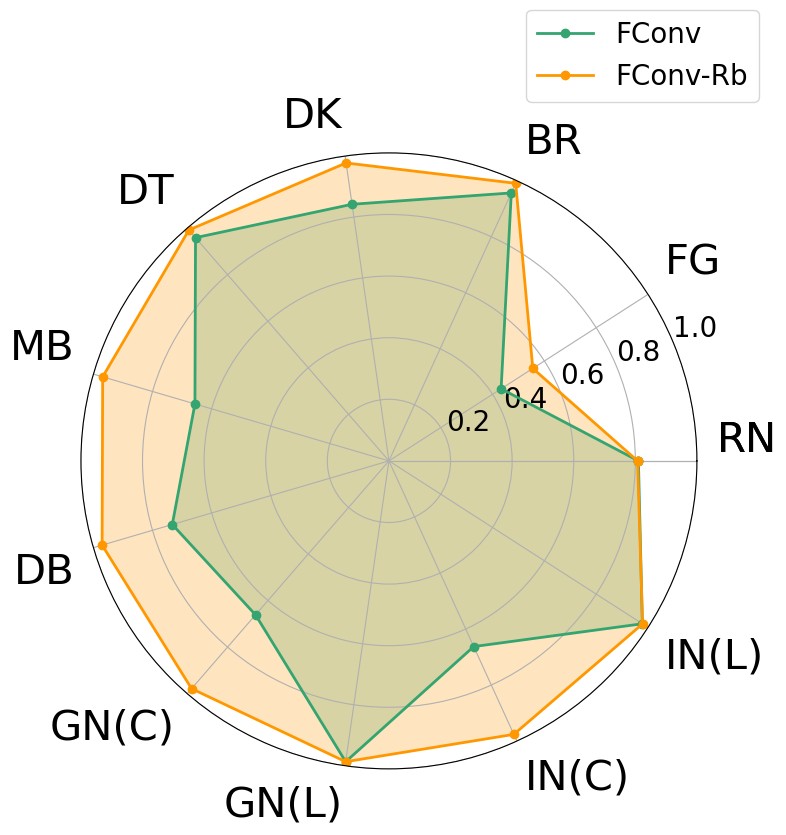}
  }
  \caption{Performance of the original and enhanced MSF.}
  
  \label{fig: rq4 rob improve}
\end{figure}

To evaluate the effectiveness of improved fusion mechanisms, we choose CLOCs and FConv as late and weak fusion systems and conduct the same experiments in RQ1 and RQ3. 
Fig.~\ref{fig: rq4 rob improve} shows the performance against corruptions of original MSF and enhanced MSF. We find that the enhanced MSF systems are significantly more robust against common corruption patterns. Furthermore, Table~\ref{tab: rq4_sig_mis} shows the improved performance ($\tilde{Rb} - Rb$, where $\tilde{Rb}$ and $Rb$ are robustness score with/without improved fusion mechanisms, respectively) against signal loss.
We find that enhanced CLOCs (CLOCs-Rb) and FConv (FConv-Rb) show promising robustness performance against partial and even complete image signal loss. For instance, when the camera signal is completely lost (100\% in Table~\ref{tab: rq4_sig_mis}), the proposed robustness enhancement strategy almost fully recovers the MSF systems' performance (highlighted in red in Table~\ref{tab: rq4_sig_mis}).


\begin{table}[t]
 \caption{
Improved performance of CLOCs-Rb and FConv-Rb against partial or complete signal loss.}
\renewcommand{\arraystretch}{1.2}
\small
 \label{tab: rq4_sig_mis}
\begin{tabular}{cccccccc}
\toprule
\multicolumn{1}{l}{\textbf{Systems}} & \textbf{Modality} & 10\%    & 25\%    & 50\%   & 75\%    & 100\%  & Avg  \\
\midrule
\multirow{2}{*}{\textbf{CLOCs-Rb}}   & C                 & -0.01 & 0.07  & 0.41 & 0.86  & \cellcolor{tab_red}0.94 & 0.45 \\
                                     & L                 & -0.01 & -0.01 & 0.00 & -0.01 & 0    & 0.00 \\
\midrule
\multirow{2}{*}{\textbf{FConv-Rb}}   & C                 & 0.00  & 0.10  & 0.52 & 0.86  & \cellcolor{tab_red}0.99 & 0.49 \\
                                     & L                 & 0.00  & 0.00  & 0.00 & 0.00  & 0    & 0.00
\\
\bottomrule
\end{tabular}
\end{table}

\begin{center}
\vspace{1mm}
\fcolorbox{black}{gray!10}{\parbox{.95\linewidth}{\textbf{Answer to RQ4:}
MSF systems with the same type of fusion mechanisms may have similar robustness issues due to their inherent properties. Deep fusion performs better against some of the corruption patterns. However, weak fusion and late fusion are easier to be repaired when facing specific robustness issues.}}
\end{center}

\vspace{1mm}
\section{Discussion}

{\bf Discussions.}
According to our findings from RQ1-3, existing AI-enabled MSF systems are not robust enough. First, corrupted signals could result in significant performance degradation of AI-enabled MSF systems. 
The data-driven nature makes it challenging to train a robust MSF system that satisfies safety and reliability requirements under all conditions.
Therefore, more research on the continuous enhancement of AI-enabled MSF is needed, such as debugging and repair. 
Our findings from RQ2 also reveal that AI-enabled MSF systems are sensitive to calibration and synchronization errors. In the real world, these two types of errors commonly exist. Even well-calibrated sensors can still be misaligned due to the changes in external environments. 
To deploy a reliable AI-enabled MSF system, developers must address the calibration issues carefully.

Modular redundancy is a critical way to improve system quality and reliability~\cite{iso21448,iso26262}. By coupling multiple sensors, AI-enabled MSF systems are expected to be robust against signal loss from one specific sensor. However, our experimental results suggest that existing work usually ignores taking this into account when designing AI-enabled MSF, resulting in a lack of robustness. Thus, future work should consider designing AI-enabled MSF systems that can still be reliable with one or more sources of signal loss. 

Though existing AI-enabled MSF systems are not robust enough, we also find it possible to repair them with fusion mechanisms improvements. In Sec.~\ref{sec:rq4}, we propose a potential repairing strategy to repair weak and late fusion mechanisms. The experimental results demonstrate their effectiveness, showing that improving fusion mechanisms could be a promising research direction.

\vspace{1mm}
{\noindent \bf Future Directions.} Based on these insights, we summarize the following future directions:

\begin{itemize}[leftmargin=*]
    \item In this work, we focus on AI-enabled MSF perception systems. However, MSF can also be used in systems beyond perception and autonomous driving. Therefore, more comprehensive benchmarks and more fine-grained robustness evaluation metrics for AI-enabled MSF systems can be considered in the future.

    \item There is an urgent need for robustness enhancement techniques to continuously improve the reliability of AI-enabled MSF systems. Based on our investigation results, improving fusion mechanisms to repair MSF systems could be a promising research direction.

    \item Different fusion mechanism-based MSF systems show different robustness issues. Therefore, practical software and system engineering approaches (e.g., testing, debugging, formal analysis, and repairing) would be needed for different MSF systems.
\end{itemize}

\vspace{1mm}
{\noindent \bf Threats to Validity.} In terms of \textit{construct validity}, 
ideally, it would be highly desirable to expose to diverse and as many corruption datasets as possible, to better approximate the robustness performance of MSF systems. Besides, randomness could also affect the process of synthesizing corrupted data.
Therefore, we try our best and adopt a large-scale systematic corrupted dataset (across thirteen corruption patterns and multiple severity levels) to comprehensively measure and analyze the robustness and reliability of MSF in our benchmark. Even though, the robustness results might still not generalize to cases of more diverse types of corruption patterns that are not evaluated in this paper.
In terms of \textit{internal validity}, one potential threat is that the leveraged weather corruption may differ from real-world weather.
To mitigate this threat, 
we choose the domain-specific physical model to simulate the properties of adverse weather for different sensors.
Further, we ensure that different sensors are sensing identical environments by controlling the hyperparameters in the physical model.
In terms of \textit{external validity}, one potential threat is that our analysis results may not be generalized to other MSF systems. To mitigate this threat, we try our best to collect a diverse set of MSF systems with different perception tasks, model structures, and fusion mechanisms.


\section{Related Works}

\noindent \textbf{Multi-sensor Fusion.}
A pioneering work of AI-enabled MSF is MV3D~\cite{chen2017multi}.
MV3D takes multi-view representations 
(i.e., front-view and bird’s eye view) 
of 3D point clouds and images as input and uses a deep fusion mechanism to combine region-wise features from multiple views.
To avoid information loss in generating view through perspective projections, EPNet~\cite{huang2020epnet} proposes a LiDAR-guided Image Fusion (LI-Fusion) module 
that 
enables the interaction between the hidden features of the point cloud and image data 
to improve 
system performance. 
CLOCs~\cite{pang2020clocs} is another representative work of late fusion, which leverages geometric and semantic consistencies of 2D and 3D output candidates to produce more accurate final detection results. 
One of the early works of weak fusion is F-PointNets~\cite{qi2018frustum}, which uses 2D bounding boxes as guidance to extract frustum in the point cloud and then estimate 3D bounding boxes.
FConv~\cite{wang2019frustum} extends the F-PointNets by proposing a sliding frustums method to aggregate local point features into frustum-level feature vectors to achieve end-to-end prediction. However, few benchmarks are available to measure the robustness and reliability of these well-designed MSF systems in open environments with corrupted/misaligned sensor signals.




%

%


\vspace{1mm}
\noindent\textbf{Robustness Benchmarks.}
Several specific robustness benchmarks designed for one data modality have been proposed.
ImageNet-C~\cite{benchmark_cifar10} evaluates the robustness of image specific recognition models against several corruptions.
Cityscapes-C~\cite{michaelis2019benchmarking} extends this ImageNet-C to 2D object detection.
However, the weather corruption in ImageNet-C and Cityscapes-C is not guaranteed to respect the underlying physics of weather conditions. 
Moreover, Mirza et al.~\cite{mirza2021robustness} evaluate the performance of autonomous driving systems under image data collected in real weather conditions. However, they do not provide a benchmark of LiDAR-based sensing modules against adverse weather conditions. Inspired by ImageNet-C, ModelNet40-C~\cite{benchmark_modelnet} measures the performance of 3D point cloud recognition models. However, these corruptions can only be applied to object-level point clouds instead of open scenes. None of these existing works has focused on benchmarking MSF systems with corrupted data from multiple different modalities. Our benchmark is thus proposed to address this.


\vspace{1mm}
\noindent\textbf{MSF Testing and Attack.}
Zhong et al.~\cite{zhong2022detecting} propose an evolutionary-based search framework to detect fusion errors for advanced driver assistance systems. Our work is parallel to them, which is to establish a general benchmark rather than testing a specific system. In addition, some recent work has investigated how to attack AI-enabled MSF systems~\cite{cao2021invisible, tu2021exploring, abdelfattah2021towards, liu2021multi}. 
Cao et al.~\cite{cao2021invisible} and Tu et al.~\cite{tu2021exploring} attack all branches of MSF systems by inserting adversarial objects. Abdelfattah et al. \cite{abdelfattah2021towards} and Liu et al. \cite{liu2021multi} investigate attacks on weak fusion and deep fusion systems, respectively.
In contrast, our benchmark aims to evaluate the robustness of the MSF systems against common real-world corruptions instead of artificial adversarial objects or perturbations.

\section{Conclusion}
In this paper, we present an early public robustness benchmark of AI-enabled MSF systems, which can further be used as a fundamental evaluation and testing framework for understanding MSF systems' limitations and potential risks. We further perform large-scale robustness evaluation on seven MSF systems against different corruption patterns including \textit{corrupted signals}, \textit{sensor misalignment}, and \textit{signal loss}. Our findings reveal that existing AI-enabled MSF are usually tightly-coupled and not robust enough. Thus, we make an early attempt to enhance the MSF system's robustness by improving fusion mechanisms. Finally, we present discussions and highlight several possible future directions in order to build robust and reliable MSF systems with the emergence of AI.

\section*{Data Availability}
Our benchmark, replication packages, and detailed evaluation results are publicly available at \href{\website}{\website}~.

\section*{Acknowledgement}
We would like to thank anonymous reviewers for their constructive comments. This project was partially funded by the National Natural Science Foundation of China under Grant No.61932012, No.61832009 and No.62002158. This work was also supported in part by Canada CIFAR AI Chairs Program, the Natural Sciences and Engineering Research Council of Canada (NSERC No.RGPIN-2021-02549, No.RGPAS-2021-00034, No.DGECR-2021-00019), as well as JST-Mirai Program Grant No.JPMJMI20B8, JSPS KAKENHI Grant No.JP20H04168, No.JP21H04877.

\balance
\bibliographystyle{ACM-Reference-Format}
\bibliography{reference}

%
\end{document}